\documentclass[24pt]{article}
\usepackage{amssymb}
\usepackage{epsfig,caption,graphicx,eepic,epic}
\usepackage{pstricks}
\usepackage{amssymb,amsmath}
\usepackage{multido}
\usepackage{array}
\begin{document}
\def\be{\begin{equation}}
\def\ee{\end{equation}}
\def\ba{\begin{eqnarray}} 
\def\ea{\end{eqnarray}}
\def\nn{\nonumber}
\newcommand{\bbf}{\mathbf}   
\newcommand{\rrm}{\mathrm}
\title{\bf Entropy and applications of the area law in open quantum systems}
\author{Tarek Khalil$^{a,}$
\footnote{E-mail address: tkhalil@ul.edu.lb}\\ 
and\\
Jean Richert$^{b}$
\footnote{E-mail address: j.mc.richert@gmail.com}\\ 
$^{a}$ Department of Physics, Faculty of Sciences(V),\\
Lebanese University, Nabatieh,
Lebanon\\ 
$^{b}$ Institut de Physique, Universit\'e de Strasbourg,\\
3, rue de l'Universit\'e, 67084 Strasbourg Cedex,\\      
France} 
\date{\today}
\maketitle 
\begin{abstract}
Open quantum systems interact with their environment and their dynamical behaviour depends strongly both on the spectral properties  of the environment and the structure of the interaction between the physical system and the environment.  We examine the consequences of these spectral and structural properties on simple but general systems in the case of deterministic (non stochastic) interactions of arbitrary  strength. The central point of interest concerns the role played by the semi-group property of the interaction in its relation with entropy and area laws properties of the system of interest.  
\end{abstract}

\maketitle  



\section{Introduction}

The interaction between an open quantum system and its environment generates a response of the system which is due to the coupling between the two parts. An important amount of work concerning the evolution of such systems has been developed over several decades. For a compilation  and recent developments concerning Markov processes one may consult f.i. refs.~\cite{riv1,pol}.\\

The evolution of an open quantum system depends strongly on the nature of the interaction, either  stochastic or deterministic, and the possible structure of the interaction. Different aspects concerning  the behaviour of the system in the latter case have been studied recently see f.i. ~\cite{kr2,kr,kr1}. There it came out that the structure of the interaction has indeed strong consequences concerning time delays, coherence, and energy transfer between the system  and its environment.\\

In the  present work we examine the evolution of the entropy. The interest in this concept has been raised in conjunction with the question of entanglement. The concept of entanglement entropy ~\cite{eis} says that the entropy of the reduced state of a subsystem grows proportionally to the boundary of the system and not proportionally to its volume as a  priori expected, following a so called "area law". This point has led to a conjecture stated by Kitaev which has been proven some time ago  ~\cite{aco}. It is the aim of the present work to examine the entropy content and evolution an open quantum in order to confirm this formal result and comment it for different types of systems in deterministic interaction with their environment.\\

The presentation of the results goes as follows. In section 2 we recall the proven conjecture. In section 3 we  introduce the formal expression and properties of the density operators which govern the total system and the  subsystem of interest. Sections 4 and 5 specializes to systems with specific dimensions, their entropy and the time evolution of the entropy. We comment the results and their consequences in section 6. Technical details concerning the calculations are shifted to appendices.
  
\section{Entropy of time dependent open quantum systems}
 
We consider a bipartite quantum system $A$ and $E$ in which $A$ is the part of physical interest  and $E$ an environment. The two subsystems are entangled and the entanglement is generated by an  interaction which works between the two parts. The study of these systems led to the concept of entanglement entropy and area laws ~\cite{eis}. In case of a dynamical evolution the maximum entanglement increase of such systems at time $t=0$ has been conjectured by Kitaev and proven later to be verified  ~\cite{bra,aco}

\ba
\Gamma_{max}=\frac{dS_{A}(t)}{dt}|_{t=0}\leq c \|\hat H_{AE}\|\log \delta
\label{eq1} 
\ea
where $\delta=min(d_{A}, d_{E})$, the smallest dimension of $A$ and $E$ space, $\|\hat H_{AE}\|$ is the norm of the interaction Hamiltonian $\hat H_{AE}$ and $c$ a constant of the order of unity. 
The entropy related to $A$ is given by 
 
\ba
S_{A}(t)=-Tr_{E}[\hat \rho_{AE}(t) \ln \hat \rho_{AE}(t)]
\label{eq2} 
\ea
where $\hat \rho_{AE}(t)$ is the density operator in $A \oplus E$ space.

In the following we aim to examine the behaviour of the entropy $S_{A}(t)$ of specific open quantum systems over finite time  intervals $[0, \tau]$ and test relation (1).

\section{The density operator of an open quentum system: general definition, divisibility property}
 
\subsection{Definition}

The general expression of the matrix elements of the density operator $\hat \rho_{AE}(\tau)$ in 
$(A \oplus E)$ space at time $\tau$ can be written as

\ba
\rho^{j_{1}j_{2}}_{\nu_{1}\nu_{2}}(\tau)=\sum_{i_{1}i_{2}}\sum_{\alpha_{1}\alpha_{2}}a_{i_{1}\alpha_{1}}
a^{*}_{i_{2}\alpha_{2}}\langle j_{1}\nu_{1}|\hat U(\tau)|i_{1}\alpha_{1} \rangle
\langle i_{2}\alpha_{2}|\hat U^{*}(\tau)|j_{2}\nu_{2} \rangle
\label{eq3} 
\ea
where $\hat U(\tau)=e^{-i\hat H \tau}$ and $\hat H= \hat H_{A}+ \hat H_{E}+ \hat H_{AE} $ is the total Hamiltonian of the subsystems ($A$, $E$)  and their interaction, and $a_{i_{k}\alpha_{k}}$ the  amplitudes of the states $|i_{k} \rangle$ and $|\alpha_{k}\rangle$ of $A \oplus E$  at time 
$t=0$. Below we shall consider that $\hat H_{AE}$ is a non stochastic interaction.

By definition $A$ and $E$ are entangled over the time interval $t=[0,\tau]$. The projected  density operator is obtained by taking the trace over the states $|\gamma\rangle$ in E space 

\ba
\rho^{j_{1}j_{2}}_{A}(\tau)=
\sum_{i_{1}i_{2} \gamma} \sum_{\alpha_{1}\alpha_{2}}a_{i_{1}\alpha_{1}}
a^{*}_{i_{2}\alpha_{2}}\langle j_{1}\gamma|\hat U(\tau)|i_{1}\alpha_{1} \rangle
\langle i_{2}\alpha_{2}|\hat U^{*}(\tau)|j_{2}\gamma\rangle
\label{eq4} 
\ea

\subsection{Divisibility property concerning the time evolution of the system}

It has been shown in ref.~\cite{kr} that the time propagator $U(\tau)$ is divisible, i.e. possesses a semi-group property in time under one of the following conditions:

\begin{itemize}

\item There is a unique state $|\gamma \rangle$ in $E$ space. This may be the case if f.i. the system $E$ is in its ground  state, see Appendix A.

\item More generally the symmetry properties of $\hat H_{E}$ and $\hat H_{AE}$ are such that 
$[\hat H_{E},\hat H_{AE}]=0$. In this case the environment $E$ stays at $t=\tau$ in one of the states it occupied at $t=0$, in practice $E$ space it is reduced to a single state during the dynamical evolution of $A \oplus E$, see Appendix B. 

\end{itemize}

\section{Two-dimensional system $A$ - one dimensional environment $E$}
 
We proceed now with the determination of the behaviour of the system $A$, the entropy and its time derivative in the restricted case where $dim_{A}=2$ and $dim_{E}=1$.
 
\subsection{The density matrix in A+E space}
 
The expression of the matrix elements of $\hat\rho_{AE}(t)$ reads

\ba
\rho^{j_{1}j_{2}\gamma}_{AE}(t)=\sum_{i_{1}i_{2}}h^{j_{1}i_{1}\gamma}(t)
h^{*j_{2}i_{2}\gamma}(t)
\label{eq5} 
\ea
where

\ba
h^{j_{1}i_{1}\gamma}(t)=a_{i_{1}\gamma}
\langle j_{1}\gamma|e^{-i\hat H t}|i_{1}\gamma \rangle
\notag\\
h^{*j_{2}i_{2}\gamma}(t)=a^{*}_{i_{2}\gamma}
\langle i_{2}\gamma|e^{+i\hat H t}|j_{2}\gamma \rangle
\label{eq6} 
\ea

\subsection{The entropy for $dim_{E}=1$, $[\hat H_{E},\hat H_{AE}]=0$ at t=0}
 
In order to obtain the entropy and its derivative we diagonalize the density matrix. The eigenvalues read

\ba
\sigma^{11}(t)=(\rho^{j_{1}j_{1}\gamma}_{AE}(t)+
\rho^{j_{2}j_{2}\gamma}_{AE}(t)-\Delta^{1/2})/2
\notag\\
 \sigma^{22}(t)=(\rho^{j_{1}j_{1}\gamma}_{AE}(t)+
\rho^{j_{2}j_{2}\gamma}_{AE}(t)+\Delta^{1/2})/2
\label{eq7}
\ea
and

\ba
\Delta=(\rho^{j_{1}j_{1}\gamma}_{AE}(t)-\rho^{j_{2}j_{2}\gamma}_{AE}(t))^{2}+
4\rho^{j_{1}j_{2}\gamma}_{AE}(t)\rho^{j_{2}j_{1}\gamma}_{AE}(t)
\label{eq8} 
\ea 
 
At time $t=0$ one finds

\ba
\sigma^{11}(0)=0
\notag\\
\sigma^{22}(0)=|a_{j_{1}\gamma}|^{2}+|a_{j_{2}\gamma}|^{2}=1
\label{eq9} 
\ea 
if the states in $A$ space are normalized.

Since it is a scalar quantity the entropy is the same in the original and diagonalized basis of
states. Hence the entropy  

\ba
S_{A}(0)=-Tr_{E}Tr_{A}S^{(d)}(0)=-\sum_{i=1,2}\sigma^{ii}(0)\ln\sigma^{ii}(0)=0
\label{eq10} 
\ea

\subsection{The entropy for $dim_{E}=1$, $[\hat H_{E},\hat H_{AE}]=0$ at t=$\tau$}

The same calculation can be also be performed at $t=\tau$. If  $dim_{E}=1$ $\hat H$ is diagonal in the basis of states in which $\hat H_{A}$  and $\hat H_{E}$ are diagonal. Indeed if  
$[\hat H_{E},\hat H_{AE}]=0$ one can factorize the evolution operator $\hat U(\tau)$
 
\ba
\hat U(\tau)=e^{-i\tau \tilde H_{A}}e^{-i\tau\hat H_{E}}=\hat U^{A}(t)e^{-i\tau\hat H_{E}}
\label{eq11} 
\ea
The matrix elements of $\tilde H_{A}$=$\hat H_{A}+\hat H_{AE}$ in $A$ space and $\hat H_{E}$ are then of the form
 
\ba
\langle j \gamma|\tilde H_{A} |k \gamma \rangle=e_{j}\delta_{jk}+\delta e_{jk}(\gamma)
\notag\\
e_{j}=\langle j|\hat H_{A} |j \rangle
\notag\\
\delta e_{jk}(\gamma)=\langle j \gamma|\hat H_{AE} |k \gamma \rangle
\notag\\
\langle j \gamma|\hat H_{E} |j \gamma \rangle=\eta_{\gamma}
\label{eq12} 
\ea
Since the entropy $S^{(d)}(0)$ is calculated in the space $A$ in which it is diagonal the time translation unitary operator $\hat U^{A}(t)$ has to be rotated into the same basis of states. The corresponding unitary operator is called $\hat V^{A}(t)$. As a consequence 

\ba
S_{A}(\tau)=Tr_{A}[ \hat V^{A\dagger}(\tau) S^{(d)}(0)\hat V^{A}(\tau)]= S_{A}(0)=0                                                           
\label{eq13} 
\ea 

The result shows that the entropy comes out to be constant over a finite interval of time in both cases of interest.

\subsection{Derivative of the entropy for $dim_{E}=1$, $[\hat H_{E},\hat H_{AE}]=0$ at $t=0$} 
 
There remains now to verify whether these results agree with the area law at $t=0$ given by 
Eq.(1). 

The expression of the entropy of the subsystem $A$ in the basis of states in which $A$  is diagonal leads to the derivative

\ba
\frac{dS_{A}^{(d)}(t)}{dt}|_{t=0}=[\frac{d\sigma^{11}(t)}{dt}(1+\ln \sigma^{11}(t))+
\frac{d\sigma^{22}(t)}{dt}(1+\ln \sigma^{22}(t))]|_{t=0}
\label{eq14}
\ea
 
The derivatives of $\sigma^{ll}(t)$ are easily calculated with some algebra starting from the expressions given by Eq.(6) at time $t=0$.
 
At $t=0$ $\sigma^{11}(0)=0$ and this is also the case for its derivative and the derivative of $\sigma^{22}(t)$. This leads to  

\ba
\frac{dS_{A}^{(d)}(t)}{dt}|_{t=0}=[\frac{d\sigma^{22}(t)}{dt}(1+\ln \sigma^{22}(t))]|_{t=0}=0
\label{eq15} 
\ea
which is in agreement with Eq.(1) since $\ln \delta=0$ in this expression. The entropy stays constant and equal to zero over finite intervals of time hence we may conclude that its derivative stays equal to zero at any time $\tau$.

\section{The entropy for $dim_{E} = N \neq 1$, $[\hat H_{A},\hat H_{AE}]=0$}

Here we consider the case where the Hamiltonian of the system $A$ commutes with the interaction  
$\hat H_{AE}$ and $dim_{E} \geq 2$. We rely on a model in which these conditions are realized and work out the entropy of the system $A$ and its derivative.

Consider the system  and its environment described by the Hamiltonian 
  
\ba
\hat H= \hat H_{A}+\hat H_{E}+\hat H_{AE}
\label{eq16}
\ea

with 

\begin{center}
\ba
\hat H_{A}=\omega \hat J_{z} 
\notag \\
\hat H_{E}=\beta b^{+}b
\notag \\
\hat H_{AE}=\eta(b^{+}+b) \hat J^{2}
\label{eq17}
\ea 
\end{center}  
where  $b^{+},b$ are boson operators, $\omega$ is the rotation frequency of the system, $\beta$ the quantum of energy of the oscillator and $\eta$ the strength parameter in the coupling interaction between $A$ and $E$.

Since $\hat J_{z}$ and $\hat J^{2}$ commute in the basis of states $\{|j m\rangle \}$ the matrix elements of $\hat H$ in $A$ space read
  
\ba
\langle jm|\hat H|jm \rangle=\omega m+\beta b^{+}b+\eta j(j+1)(b^{+}+b) 
\label{eq18}
\ea 
  
The expression of the density operator $\hat \rho_{A}(t)$ at time $t$ is then obtained by taking the trace over the environment states of the total Hamiltonian $\hat \rho(t)$ leading to 

\ba
\hat \rho_{A}(t)=Tr_{E}\hat \rho(t) 
\label{eq19}
\ea 
whose matrix elements read  
  
\ba
\rho^{j m_{1}, j m_{2}}_{A}(t)=\rho^{j m_{1}, j m_{2}}_{0}(t)\Omega_{E}(j,j,t)
\label{eq20}
\ea 
with  
  
\ba
\rho^{j m_{1}, j m_{2}}_{0}(t)=e^{[-i\omega(m_{1}-m_{2})]t}/(2j+1)
\label{eq21} 
\ea  
The bosonic environment contribution can be put in the following form 
  
\ba
\Omega_{E}(j,j,t)=\sum_{n=0}^{N}\frac{1}{n!}\sum_{n',n^{"}}\frac{E_{n,n'}(j,t)
E^{*}_{n^{"},n}(j_{2},t)}{[(n'!)(n''!)]^{1/2}}
\label{eq22} 
\ea  
The expression  is exact. The Zassenhaus development formulated in Appendix D was used in order to work out the $\Omega_{E}$ ~\cite{za}. The expressions of the polynomials $E_{n,n'}(t)$ and 
$E^{*}_{n'',n}(t)$ are developed in Appendix E.

The entropy $S_{A}(t)= -Tr_{E}\hat \rho(t) ln\hat \rho(t)$ can be worked out analytically. For $j=1/2$
and in the frame in which $\rho^{j m_{1}, j m_{2}}_{0}(t)$ is diagonal it reads

\ba
S_{A}(t)=-\Omega_{E}(j,j,t)\ln \Omega_{E}(j,j,t)
\label{23}
\ea
As an example we restrict the bosonic space to $N=1$, i.e. a subsystem $E$ of dimension 2

\ba
\Omega_{E}(j,j,t)=\Pi(t)e^{2\Psi(t)}
\label{24}
\ea
The algebraic expression of $\Omega_{E}$ is given by 

\ba
\Pi(t)=2[1+\alpha(t)(\zeta^{2}(t)+\zeta(t)-1)+2\alpha^{2}(t)(1-\zeta(t))]
\notag\\
+\zeta(t)[\zeta^{3}(t)+\zeta^{2}(t)+\zeta(t)-1]+\alpha^{4}(t)(\zeta^{4}(t)-1)
\label{25}
\ea 
using the expressions given in Appendix E.

The calculations of the entropy at $t=0$ leads to 

\ba
S_{A}(t)|_{t=0}= -2\ln 2
\label{26}
\ea 
and shows an intricate oscillatory behaviour for $t\neq 0$ which  may be periodic in time or not depending on the commensurability of the oscillating functions $\alpha(t)$ and $\zeta(t)$. The  derivative of the  entropy can be worked out in a similar manner

\ba
dS_{A}(t)/dt|_{t=0}= -\gamma(\ln2+1)
\label{27}
\ea 
which can be rewritten as

\ba
\Gamma=dS_{A}(t)/dt|_{t=0}= -3/4 (1+1/\ln2) \eta \ln2
\label{28}
\ea 
since $j=1/2$, $\gamma=j(j+1)\eta$, and $\eta$ is the strength of the interaction Hamiltonian. One recognizes here the expression $\Gamma_{max}$ of the Kitaev conjecture shown in Eq.1.~\cite{bra,aco}. 
If $\eta$ is positive $\Gamma$ is negative, hence smaller than $\Gamma_{max}$. If $\eta$ is negative
the equal sign shows that $\Gamma=\Gamma_{max}$ in Eq.28. Introducing $c=3/4 (1+1/\ln2)$  shows that 
this constant is close to $c=1$ and one may conjecture that it decreases with increasing $dim_{d}$ so that $c(d)$ in Eq.1 decreases as expected by the theory, see ~\cite{bra}.

The dimension of the  bosonic environment space can evidently be extended to any dimension $N$.

\section{Comments and physical interpretation of the results, conclusion}

The present results lead to a certain number of comments:

\begin{itemize}

\item The present process for $dim_{E}=1$ could be experimentally realized if the spectrum of subsystem $E$ could be reduced to the ground state  at very low temperature or if the coupling Hamiltonian between the two subsystems would commute with the Hamiltonian of the bath. One may however notice that both processes may be difficult to implement experimentally.

\item The present results were restricted to $dim_{A}=2$. They can be generalized to any dimension of $A$ space. The behaviour of the entropy and its derivative at $t=0$ should not change. This is expected because the restriction to a one-dimensional $E$ space forbids any energy exchange between the system and its environment keeping the entropy constant.  
     
\item Generalization to higher dimensions in $E$ space: if $[\hat H_{E},\hat H_{AE}]=0$ the subsystem $E$ stays in the state in which it lived at time $t=0$, see Appendix B. If however the  states in $A$ and $E$ space are entangled at time $t=0$ this is not necessarily  the case. Then the subsystem $E$ generates memory effects which can be interpreted as due to "jumps" from one state to the other and this should generate entropy in subsystem $A$, see Appendix C. It is expected that the entropy changes over a finite interval of time.    

\item Last the constancy of $S_{A}(\tau)$, equal to zero or finite, may be related to reversibility of the dynamical  process~\cite{zwa1,zwa2}. There is no energy exchange between the $A$ and $E$ subsystem in this case even though energy may be stored in $\hat H_{AE}$.

\item  In the case where $[\hat H_{A},\hat H_{AE}]=0$ the model shows a very different
behaviour of the entropy of the subsystem $A$ in section $5$, independently of the size of $E$ space. There the entropy is finite and reversibility does no longer survive. The results are in agreement with those predicted by the area law.

\item The model introduced in this case was restricted to a more or less realistic physical case corresponding to the description of a non decohering system. The introduction of a more realistic coupling in which the operator $\hat J^{2}$ would f.i. be replaced by $\hat J_{z}$ would however not change the qualitative behaviour of the entropy. 

\item The general case corresponding to $[\hat H_{A},\hat H_{AE}]\neq 0$ and 
$[\hat H_{E},\hat H_{AE}]\neq 0$ leads to a strongly model dependent behaviour.

\end{itemize}

In the present work we examined the properties of some specific tractable types of open quantum systems concerning their evolution with time and questioned the prediction of area laws which
are related to entanglement entropy. We worked out these quantities on simple physical systems and examined explicitly the connection between these concepts. We showed in particular the role played by the structure of the environment of the examined  system and the crucial importance of the properties of the interaction acting between the system and its environment.  

For reasons of completness we repeat in the following appendices results which have already been 
shown in former work ~\cite{kr4}.

\section{Appendix A: Divisibility when $dim_{E}=1$}

Here we consider the case where an open system possesses the divisibility property. Its evolution is described by a density operator  $\hat\rho_{A}(t)$ which evolves in time under the action of the evolution operator $\hat T(t,0)$

\ba
\hat\rho_{A}(t)=\hat T(t,0)\hat\rho_{A}(0)
\label{eq29}
\ea

At time $t>0$ the reduced density operator in $A$ space is $\hat\rho_{A}(t)=Tr_{E}[\hat\rho(t)]$ where $\hat\rho(t)$ is the density operator of the total system $S+E$. Under the ssumption that subsystems $A$ and $E$ do not interact at  time $t=0$ it can be written ~\cite{vlb}

\ba
\hat\rho_{A}(t)=\sum_{i_{1},i_{2}}c_{i_{1}}c_{i_{2}}^{*}\hat\Phi_{i_{1},i_{2}}(t,0)
\label{eq30}
\ea
with

\ba
\hat\Phi_{i_{1},i_{2}}(t,0)=\sum_{j_{1},j_{2}}C_{(i_{1},i_{2}),(j_{1},j_{2})}(t,0)|j_{1}\rangle_{S}\langle j_{2}|
\label{eq31}
\ea

where the super matrix $C$ reads

\ba
C_{(i_{1},i_{2}),(j_{1},j_{2})}(t,0)=\sum_{\alpha_{1},\alpha_{2},\gamma}d_{\alpha_{1},\alpha_{2}}
U_{(i_{1}j_{1}),(\alpha_{1}\gamma)}(t,0) U_{(i_{2}j_{2}),(\alpha_{2}\gamma)}^{*}(t,0)
\label{eq32}
\ea
where $ d_{\alpha_{1},\alpha_{2}}$ is the weight attached to the  states $\alpha_{1}$ and $\alpha_{2}$
and

\ba
U_{(i_{1}j_{1}),(\alpha_{1}\gamma)}(t,0)=\langle j_{1}\gamma|\hat U(t,0)|i_{1} \alpha_{1}\rangle
\notag\\
U^{*}_{(i_{2}j_{2}),(\alpha_{2}\gamma)}(t,0)=\langle i_{2} \alpha_{2}|\hat U^{*}(t,0)|j_{2} \gamma \rangle
\label{eq33}
\ea

The divisibility criterion reads

\ba
\hat\rho_{A}(t,0)=\hat T(t,\tau)\hat T(\tau,0)\hat\rho_{A}(0)
\label{eq34}
\ea
with $\tau$ in the interval $[0,t]$.

For this to be realized the following relation must be verified by the super matrix $C$

\ba
C_{(i_{1},i_{2}),(k_{1},k_{2})}(t,0)= \sum_{j_{1},j_{2}}C_{(i_{1},i_{2}),(j_{1},j_{2})}(t_{s},0)
C_{(j_{1},j_{2}),(k_{1},k_{2})}(t,t_{s})
\label{eq35}
\ea 

Using the explicit expression of the super matrix $C$ given by Eqs.(30-31) the divisibility constraint in Eq.(33) for fixed states $(i_{1}, i_{2})$, $(k_{1}, k_{2})$ imposes the following relation 
 
\ba
\sum_{\alpha_{1},\alpha_{2},\gamma}d_{\alpha_{1},\alpha_{2}}U_{(i_{1}k_{1}),(\alpha_{1}\gamma)}(t)
U^{*}_{(i_{2}k_{2}),(\alpha_{2}\gamma)}(t)=
\sum_{j_{1},j_{2}}\sum_{\alpha_{1},\alpha_{2},\beta_{1},\beta_{2}}d_{\alpha_{1},\alpha_{2}} 
d_{\beta_{1},\beta_{2}}
\notag \\
\sum_{\gamma,\delta}U_{(j_{1}k_{1}),(\beta_{1}\delta)}(t-t_{s})U_{(i_{1}j_{1}),(\alpha_{1}\gamma)}(t_{s})
U^{*}_{(j_{2}k_{2}),(\beta_{2}\delta)}(t-t_{s})U^{*}_{(i_{2}j_{2}),(\alpha_{2}\gamma)}(t_{s}) 
\label{eq36}
\ea

In order to find a solution to this equality and without loss of generality we consider the case where the density matrix in $E$ space is diagonal. Then the equality reads 
 
\ba
\sum_{\alpha, \gamma}d_{\alpha,\alpha}U_{(i_{1}k_{1}),(\alpha \gamma)}(t)
U^{*}_{(i_{2}k_{2}),(\alpha \gamma)}(t)=
\sum_{j_{1},j_{2}}\sum_{\alpha,\beta}d_{\alpha,\alpha} d_{\beta,\beta}
\notag \\
\sum_{\gamma,\delta}U_{(j_{1}k_{1}),(\beta \delta)}(t-t_{s})U_{(i_{1}j_{1}),(\alpha \gamma)}(t_{s})
U^{*}_{(j_{2}k_{2}),(\beta \delta)}(t-t_{s})U^{*}_{(i_{2}j_{2}),(\alpha \gamma)}(t_{s}) 
\label{eq37}
\ea

A sufficient condition to realize the equality is obtained if $d_{\beta,\beta}=d_{\alpha,\alpha}$ and consequently if the weights $d$ on both sides are to be the same one ends up with $d_{\alpha,\alpha}=1$. This last condition imposes a unique state in $E$ space, say $|\eta \rangle$. In this case $d_{\eta,\eta}=1$ and Eq.(36) reduces to  
 
\ba
U_{(i_{1}k_{1}),(\eta \eta)}(t)U^{*}_{(i_{2}k_{2}),(\eta \eta)}(t)=
\sum_{j_{1}} U_{(i_{1}j_{1}),(\eta \eta)}(t_{s})U_{(j_{1}k_{1}),(\eta \eta)}(t-t_{s})
\notag\\
\sum_{j_{2}} U^{*}_{(j_{2}k_{2}),(\eta \eta)}(t-t_{s})U^{*}_{(i_{2}j_{2}),(\eta \eta)}(t_{s}) 
\label{eq38}
\ea
which proves the equality.

The interaction Hamiltonian $\hat H_{AE}$ generates entanglement between the system $A$ and the environment $E$. This coupling is also the source of time retardation memory effects in the time behaviour of the system $A$. One may ask how the absence of retardation imposed by the strict divisibility constraint is correlated with the entanglement induced by the coupling between the two systems.

When divisibility is strictly verified by means of the sufficient condition found above the matrix elements of $\hat\rho(t)$ take the form

\ba
\rho_{A}^{j_{1},j_{2}}(t)=\sum_{i_{1},i_{2}}a_{i_{1},\eta} a_{i_{2},\eta}^{*}
\langle j_{1}\eta|\hat U(t)|i_{1} \eta\rangle 
\langle i_{2} \eta|\hat U^{*}(t)|j_{2} \eta\rangle |j_{1}\rangle \langle j_{2}| 
\label{eq39} 
\ea
In this case considered in the text entanglement between $A$ and $E$ is reduced to the coupling of the system to a one-dimensional environment space. The Hilbert space of the total system $A+E$ reduces in practice to dimension $d+1$ where $d$ is the dimension of $A$.\\ 
 
\section{Appendix B: Divisibility when $[\hat H_{E},\hat H_{AE}]=0$ }

We show now that the semi-group (divisibility) ptoperty  can be realized even if there is more than one state in $E$ space. To see this we introduce the explicit expression of the master equation which governs an open quantum system in a time local regime.  Its expression reads~\cite{gl,gk,lain,hall,pearl}
\ba
\frac{d}{dt}\hat\rho_{A}(t)=\sum_{n}\hat L_{n}\hat\rho_{A}(t)\hat R_{n}^{+}
\label{eq27}
\ea
where $\hat L_{n}$ and $\hat R_{n}$ are time independent operators.\\

Using the general form of the density operator $\hat\rho_{S}(t)$ given by Eqs. (29-30) 

\ba
\hat\rho_{A}^{j_{1}j_{2}}(t)=\sum_{i_{1}i_{2}}c_{i_{1}}c^{*}_{i_{2}}\sum_{\alpha\alpha_{2}, \gamma}d_{\alpha_{1},\alpha_{2}}\langle j_{1}\gamma|\hat U(t,t_{0})|i_{1} \alpha_{1}\rangle
\notag\\
\langle i_{2} \alpha_{2}|\hat U^{*}(t,t_{0})|j_{2} \gamma \rangle
\label{eq28}
\ea
and taking its time derivative leads to two contributions to the matrix elements of the operator 

\ba
\frac{d}{dt}\rho_{A1}^{j_{1}j_{2}}(t)=(-i)\sum_{i_{1}i_{2}}c_{i_{1}}c^{*}_{i_{2}}\sum_{\alpha_{1},\alpha_{2}}d_{\alpha_{1},\alpha_{2}}
\sum_{\beta \gamma k_{1}}\langle j_{1}\gamma|\hat H|k_{1} \beta \rangle
\notag\\
\langle k_{1}\beta|e^{-i\hat H(t-t_{0})}|i_{1} \alpha_{1} \rangle \langle i_{2}\alpha_{2}|e^{i\hat H(t-t_{0})}|j_{2} \gamma \rangle 
\notag\\
\frac{d}{dt}\rho_{A2}^{j_{1}j_{2}}(t)=(+i)\sum_{i_{1}i_{2}}c_{i_{1}}c^{*}_{i_{2}}\sum_{\alpha_{1},\alpha_{2}}d_{\alpha_{1},\alpha_{2}}
\sum_{\beta \gamma k_{2}}\langle j_{1}\gamma|e^{-i\hat H(t-t_{0})}|i_{1} \alpha_{1} \rangle  
\notag\\
\langle i_{2}\alpha_{2}|e^{i\hat H(t-t_{0})}|k_{2} \beta \rangle \langle k_{2}\beta|\hat H|j_{2} \gamma \rangle 
\label{eq29}
\ea
and                  

\ba
\frac{d}{dt}\hat\rho_{A}^{j_{1}j_{2}}(t)=\frac{d}{dt}[\rho_{A1}^{j_{1}j_{2}}(t) +\rho_{A2}^{j_{1}j_{2}}(t)]
\label{eq30}
\ea
From the explicit expression of the density operator matrix element given by Eqs. (41-43) one sees that the structure of the master equation given by Eq.(40) which induces the divisibility can only be realized if $|\beta \rangle=|\gamma \rangle$. Two general solutions can be found:

\begin{itemize}
\item There is only one state $|\gamma \rangle$ in $E$ space. This result has already been seen on the expression of the density operator above(see Appendix A).

\item  If the environment stays in a fixed state $|\gamma \rangle$, i.e. if the Hamiltonian 
$\tilde H=\hat H_{E}+\hat H_{AE}$ is diagonal in a basis of states in which $\hat H_{E}$ is diagonal. Then, if the system starts in a given state $|\gamma\rangle$ it will stay in this state over the whole interval of time and the density operator will be characterized by a definite index $\gamma$,  
$\hat\rho_{A \gamma}(t)$. The central  point to notice here is the fact that this happens if 
$[\hat H_{E},\hat  H_{AE}]$=0. The result is also valid if the subsystems $A$ and $E$ are entangled from the start at $t=0$. 

\end{itemize}

\section{Appendix C: Memory effects for $dim_{E} \neq 1$}

\begin{itemize}

\item Memory effects and absence of divisibility: two-time approach

We use the projection formalism ~\cite{nak,ck1,zwa,veg} and the expression developed in Appendix A in order to analyze the time evolution of the density operator of the total system $A+E$ 

\ba
\hat \rho(t,t_{0})=\sum_{i_{1},i_{2}}c_{i_{1}}c^{*}_{i_{2}}\sum_{\alpha}d_{\alpha \alpha}U(t,t_{0})
|i_{1} \alpha \rangle \langle i_{2} \alpha| U^{+}(t,t_{0})
\label{eq31}
\ea
We write the expression of $\hat \rho(t,t_{0})$ in a basis of states in which $\hat H_{E}$ is diagonal.

We introduce projection operators $\hat P$ and $\hat Q$ in $E$ space such that

\ba
\hat P \hat \rho(t,t_{0})=\sum_{k=1}^{n}|\gamma_{k}\rangle \langle \gamma_{k}|\hat \rho(t,t_{0})
\notag\\
\hat Q \hat \rho(t,t_{0})=\sum_{l=n+1}^{N}|\gamma_{l}\rangle \langle \gamma_{l}|\hat \rho(t,t_{0})
\label{eq32}
\ea
where $N$ is the total finite or infinite number of states in $E$ space and $\hat P+\hat Q=\hat I$ 
where $\hat I$ is the identity operator.

The evolution of the density operator is given the Liouvillian equation
\ba
\frac { d \hat \rho(t,t_{0})}{ dt}=\hat L(t) \hat \rho(t,t_{0})=-i[\hat H,\hat \rho(t,t_{0})]
\label{eq33}
\ea

Projecting this equation respectively on $\hat P$ and $\hat Q$ subspaces leads to a set of two coupled  equation

\ba
\frac{d \hat P \hat \rho(t,t_{0})}{dt}= \hat P \hat L(t)\hat P \hat \rho(t,t_{0})+
\hat P \hat L(t)\hat Q \hat \rho(t,t_{0}) (a)
\notag\\
\frac{d \hat Q \hat \rho(t,t_{0})}{dt}= \hat Q \hat L(t)\hat Q \hat \rho(t,t_{0})+
\hat Q \hat L(t)\hat P \hat \rho(t,t_{0}) (b)
\label{eq34}
\ea

Choosing $t_{0}=0$ in order to simplify the equations and solving formally the second equation gives
\ba
\hat Q\hat \rho(t)=e^{\hat Q\hat L(t)t}\hat Q\hat \rho(t=0)+ \int ^{t}_{0}
dt'e^{\hat Q \hat L(t')t'}\hat Q \hat L(t')\hat P \hat \rho(t-t^{'})
\label{eq35}
\ea

If inserted into the first equation one obtains
\ba
\frac{d \hat P \hat \rho(t)}{dt}= \hat P \hat L(t)\hat P \hat \rho(t)+\hat P\hat L(t)
e^{\hat Q\hat L(t)t}\hat  Q\hat \rho(0)+\hat P \hat L(t)*
\notag\\
\int ^{t}_{0}dt'e^{\hat Q\hat L(t')t'}\hat Q \hat L(t')\hat P \hat \rho(t-t')
\label{eq36}
\ea  

This first order two-time integro-differential equation reduces to an ordinary one-time differential equation under one of the the following conditions:

\begin{itemize}

\item There is only one state $|\gamma\rangle$ in $E$ space. Then dim$\hat P=1$ and dim$\hat Q=0$.
As a consequence Eq.(46) reduces to
\ba
\frac{d \hat P \hat \rho(t)}{dt}=i\hat P[\hat P \hat \rho(t),\hat H]
\label{eq37}
\ea

\item The density operator at $t=0$ is such that $\hat P \hat \rho(0) \hat Q=0$, i.e. 
$\hat \rho(0)$ is block diagonal and furthermore $[\hat H_{E},\hat H_{AE}]=0$ in a basis of states in which $\hat H_{E}$ is diagonal. Then $\hat P \hat H \hat Q = 0$ and in the second terms of Eqs.(47a) and (47b), $ \hat P [\hat Q\hat  \rho(t),\hat H]=0$ and 
$\hat Q [\hat P \hat \rho(t),\hat H]=0$. This eliminates the second terms in Eqs.(47) which decouple.

\end{itemize}

Hence the evolution of the P-projected density operator $\hat P\hat \rho(t)$ is local in time and possesses the divisibility property. This result is again in agreement with the results obtained above and also with ref.~\cite{ck}. Finally the  evolution of the density operator in $A$  space $\hat \rho_{A}(t)$ is governed by 

\ba
Tr_{PE}\frac{d \hat P \hat \rho(t)}{dt}=iTr_{PE}\hat P[\hat P \hat \rho(t),\tilde H]
\notag
\ea
where $PE$ stands for the $P$ projection of $E$ space and $\tilde H=\hat H_{A}+\hat H_{AE}$.

\item Memory effects and absence of divisibility: one-time approach

Consider the case $[\hat H_{E},\hat H_{AE}]\neq 0$ in a basis of states in which $\hat H_{E}$ is diagonal. The violation of divisibility is then realized because $\hat H_{AE}$ possesses non-diagonal elements. Then the evolution of the density matrix is described by a master equation whose matrix elements for a fixed state $|\gamma\rangle$ in $E$ space depends on a unique time variable and takes the form 

\ba
\frac{d \hat \rho^{ik}_{A\gamma}(t)}{dt}=(-i)[\hat H_{d}^{\gamma},\hat \rho_{A\gamma}(t)]^{ik}+
(-i)[\sum_{\beta \neq \gamma}[\Omega^{ik}_{\gamma \beta}(t)-\Omega^{ik}_{\beta \gamma}(t)]
\label{eq51}
\ea
where $\hat H_{d}^{\gamma}$ is the diagonal part in $E$ space of $\hat H$ for fixed $\gamma$ and  

\ba
\Omega^{ik}_{\gamma \beta}(t)=\sum_{j}\langle i \gamma|\hat H_{AE}|\beta j\rangle 
\langle j \beta |\hat \rho(t)|\gamma k\rangle
\notag\\
\Omega^{ik}_{\beta \gamma}(t)=\sum_{j}\langle i \gamma|\hat \rho(t|\beta j\rangle
\langle j \beta |\hat H_{AE}|\gamma k\rangle
\label{eq52}
\ea

In the present formulation the master equation depends on a unique time variable although it describes a non divisible process. Physically it is the fact for the environment to get the opportunity to "jump" from a state $|\gamma\rangle$ to another state $|\beta\rangle$ which produces necessarily a time delay. This time delay induces the violation of the semi-group property when this delay is absent in the process. Here the strength of the violation is measured by the strength of the non-diagonal elements. 
 
\end{itemize}

\section{Appendix D: the Zassenhaus development} 
 
If $X=-i(t-t_{0})(\hat H_{A}+\hat H_{E})$ and $Y=-i(t-t_{0})\hat H_{AE}$

\ba
e^{X+Y}=e^{X}\otimes e^{Y}\otimes e^{-c_{2}(X,Y)/2!}\otimes e^{-c_{3}(X,Y)/3!}\otimes e^{-c_{4}(X,Y)/4!}...
\label{eq53}
\ea

where

\begin{center}
$c_{2}(X,Y)=[X,Y]$\\ 
$c_{3}(X,Y)=2[[X,Y],Y]+[[X,Y],X]$\\ 
$c_{4}(X,Y)=3[[[X,Y],Y],Y]+3[[[X,Y],X],Y]+[[[X,Y],X],X]$, etc.\\
\end{center} 
 
The series has an infinite number of term which can be generated iteratively in a straightforward way  
~\cite{ca}. If $[X,Y]=0$ the truncation at the  third term  leads to the factorisation of the $X$ and the $Y$ contribution. If $[X,Y]=c$ where $c$ is a c-number the expression corresponds to the well-known Baker-Campbell-Hausdorff formula.\\  


 \section{Appendix E: The bosonic content of the density operator}
 
The expressions of the bosonic contributions to the density matrix $\rho^{j m_{1}, j m_{2}}_{s}(t)$ are given by 
 
\ba
E_{n,n'}(j,t)=e^{-i\beta t}\sum_{n\geq n_{2},n_{3}\geq n_{2}}\sum_{n_{3}\geq n_{4},n'\geq n_{4}}(-i)^{n+n_{3}}
(-1)^{n'+n_{2}-n_{4}}
\notag\\
\frac{n!n'!(n_{3}!)^{2}[\alpha(t)^{n+n_{3}-2n_{2}}][\zeta(t)^{n_{3}+n'-2n_{4}}]}{(n-n_{2})!(n_{3}-n_{4})!
(n_{3}-n_{2})!(n'-n_{4})!}e^{\Psi(t)}
\label{eq55}      
\ea
and

\ba
E^{*}_{n^{"},n}(t;j)=e^{i\beta t}\sum_{n^{"}\geq n_{2},n_{3}\geq n_{2}}\sum_{n_{3}\geq n_{4},n\geq n_{4}}i^{n^{"}+n_{3}}
(-1)^{n+n_{2}-n_{4}}
\notag\\
\frac{n^{"}!n!(n_{3}!)^{2}[\alpha(t)^{n^{"}+n_{3}-2n_{2}}][\zeta(t)^{n+n_{3}-2n_{4}}]}{(n^{"}-n_{2})!(n_{3}-n_{2})!(n_{3}-n_{4})!(n-n_{4})!}e^{\Psi(t)}
\label{eq56}      
\ea

The different quantities which enter these expressions are 

\ba
\alpha(t)=\frac{\gamma(j)\sin\beta t}{\beta}
\label{eq57}      
\ea

\ba
\zeta(t)=\frac{\beta[1-\cos\gamma(j)t]}{\gamma(j)}
\label{eq58}      
\ea

\ba
\gamma(j)=\eta j(j+1)
\label{eq59}      
\ea

\ba
\Psi(t)=-\frac{1}{2}[\frac{\gamma^{2}(j)\sin^{2}(\beta t)}{\beta^{2}}+\frac{\beta^{2}(1-\cos\gamma(j)t)^{2}}{\gamma^{2}(j)}]                           
\label{eq60}      
\ea

\end{document}